\documentclass{optica-article}

\journal{opticajournal} 

\articletype{Research Article}

\usepackage{lineno}

\begin{document}

\title{Integrated time-bin entangled quantum light source on a 4H-SiC microring chip}

\author{Hong Zeng,\authormark{1,2,3,$\dagger$} Bing-Cheng Yang,\authormark{4,5,$\dagger$} Yun-Ru Fan,\authormark{1,2,3,11$\dagger$} Li-Ping Zhou,\authormark{4} Cheng-Li Wang,\authormark{4,5,12} Bo-Wen Chen,\authormark{4,5} Ai-Lun Yi,\authormark{4,5} Yong Geng,\authormark{6} Guang-Wei Deng,\authormark{1,3} You Wang,\authormark{1,7} Hai-Zhi Song,\authormark{1,7} Jun-Tao Zhang,\authormark{8} Hao Li,\authormark{9} Li-Xing You,\authormark{9} Zi-Hao Zhan,\authormark{8} Kai Guo,\authormark{8,13} Xin Ou,\authormark{4,5,14} Guang-Can Guo,\authormark{1,2,3,10} and Qiang Zhou\authormark{1,2,3,10,15}}

\address{\authormark{1}Institute of Fundamental and Frontier Sciences, University of Electronic Science and Technology of China, Chengdu 611731, China\\
\authormark{2}Center for Quantum Internet, Tianfu Jiangxi Laboratory, Chengdu 641419, China\\
\authormark{3}Key Laboratory of Quantum Physics and Photonic Quantum Information, Ministry of Education, University of Electronic Science and Technology of China, Chengdu 611731, China\\
\authormark{4}State Key Laboratory of Materials for Integrated Circuits, Shanghai Institute of Microsystem and Information Technology, Chinese Academy of Sciences, Shanghai 20050, China\\
\authormark{5}Center of Materials Science and Optoelectronics Engineering, University of Chinese Academy of Sciences, Beijing 100049, China\\
\authormark{6}Key Lab of Optical Fiber Sensing and Communication Networks, University of Electronic Science and Technology of China, Chengdu 611731, China\\
\authormark{7}Southwest Institute of Technical Physics, Chengdu 610041, China\\
\authormark{8}Institute of Systems Engineering, AMS Beijing 100141, China\\
\authormark{9}National Key Laboratory of Materials for Integrated Circuits, Shanghai Institute of Microsystem and Information Technology, Chinese Academy of Sciences, Shanghai 200050, China\\
\authormark{10}CAS Center for Excellence in Quantum Information and Quantum Physics, University of Science and Technology of China, Hefei 230026, China\\
\authormark{11}yunrufan@uestc.edu.cn\\
\authormark{12}wangcl@mail.sim.ac.cn\\
\authormark{13}guokai07203@hotmail.com\\
\authormark{14}ouxin@mail.sim.ac.cn\\
\authormark{15}zhouqiang@uestc.edu.cn
}
\noindent $\dagger$These authors contributed equally to this work.


\begin{abstract*} 
Integrated time-bin-entangled photon-pair source with cavity-enhanced nonlinear optical processes is essential for quantum information technologies. 
However, microcavities with a high quality factor inherently introduce a trade-off between generation efficiency and photon bandwidth, which hinders the development of high-speed quantum networks with an integrated source. 
Here, we address this challenge by optimizing the nonlinearity property of the material and the geometry of the integrated microring resonator with a 4H-silicon carbide platform.
Operating at a loaded quality factor of $1.9\times10^{5}$ - spectral bandwidth of $1.0~\mathrm{GHz}$ and pumped with 300-ps double pulses separated by 1.25 ns at a repetition rate of $160~\mathrm{MHz}$, the device achieves a time-bin-entangled photon-pair generation rate of $1.35\times10^{7}~\mathrm{s}^{-1}\,\mathrm{mW}^{-2}$.
A raw visibility of $95.55\pm0.18\%$ is measured, showing a violation of Bell's inequality by more than 138 standard deviations, and a fidelity of $94.37\pm0.22\%$ is obtained by quantum state tomography.
These results provide a scalable pathway to an efficient and broadband time-bin entangled quantum light source, overcoming intrinsic limitations of cavity-based designs and advancing integrated platforms for future quantum communication networks.

\end{abstract*}

\section{Introduction}
 Photonic quantum entanglement is central to photonic quantum technologies, underpinning advances in quantum communication\cite{marcikic2003long, zhang2025classical}, quantum computation\cite{gottesman1999demonstrating, psiquantum2025manufacturable, aghaee2025scaling}, and quantum metrology\cite{afek2010high, cao2024multi, finkelstein2024universal}.~Various photonic degrees of freedom, including path\cite{wang2018multidimensional}, polarization\cite{kwiat2001experimental}, orbital angular momentum\cite{erhard2018twisted,krenn2014generation}, frequency bin\cite{kues2017chip}, energy–time\cite{franson1989bell}, and time-bin\cite{marcikic2002time, marcikic2004distribution, reimer2016generation,yu2025quantum} have been exploited to encode and process quantum information.~Among these, time-bin encoding has become a widely adopted approach for distributing qubits and high-dimensional quantum states, due to its robustness against environmental noise such as thermal and mechanical fluctuations, depolarization, and fiber birefringence\cite{yu2024time,singh2025photonic,xiong2015compact,zhang2018integrated}.
Integrated quantum light sources based on photonic integrated circuits have become key building blocks for scalable quantum technologies \cite{caspani2017integrated,wang2020integrated, elshaari2020hybrid, mahmudlu2023fully, wang2025scalable,  kramnik2025scalable}.~Cavity-enhanced architectures, such as microrings\cite{reimer2016generation, kues2017chip, guo2017parametric, reimer2019high, ma2020ultrabright, steiner2021ultrabright, fan2023multi, chen2024ultralow, zeng2024quantum, rahmouni2024entangled, pang2025versatile}, microdisks\cite{Luo17QiangL, Kumar20, Ren2022microdisk}, and microspheres\cite{Ortiz-Ricardo2021}, leverage high quality factors (Q-factors), small mode volumes, and engineered dispersion to significantly enhance nonlinear interactions, enabling highly efficient photon-pair generation beyond waveguide-based implementations.~For instance, bright and narrowband entangled sources have been demonstrated on an integrated photonics platform with ultralow loss quality factors exceeding $10^6$\cite{chen2024ultralow, Ortiz-Ricardo2021, ramelow2015silicon}, which are critical to an efficient photon-atom interface.~However, the narrow linewidth of high-Q resonators intrinsically limits the achievable data rate, in tension with the broadband, efficient photon-pair sources required for high-speed quantum communication networks. ~F. Samara et al. demonstrated sequential time-bin entanglement generation in a low-Q $Si_3N_4$ microring resonator with a bandwidth of 1.75 GHz and a photon-pair generation rate (PGR) of $2.5\times10^3$   $~\mathrm{s^{-1}mW^{-2}}$\cite{samara2019high}. ~Balancing the efficiency enhancement benefits of high-$Q$ resonators with the broadband operation of low-$Q$ devices remains a key challenge, originating from the conflicting constraints that efficiency and bandwidth impose on the cavity quality factor with a given material and fabrication platform \cite{xu2025quantum,chen2023pushing}.

A promising route to decoupling the efficiency–bandwidth trade-off lies in enhancing the intrinsic nonlinearity, which sustains high photon-pair generation rates even in low-Q cavities designed for broad spectral bandwidths.~This enhancement can be realized by tuning the Kerr nonlinearity via stoichiometry and doping, and by exploiting the anisotropy of the nonlinear tensor (differentiating TE and TM modes)\cite{li2023measurement, lu2024strong, cai2022octave}.~These modifications must be implemented while maintaining the pristine crystalline quality essential for suppressing the Raman noise. 4H-silicon carbide (SiC) crystalline platform, offering both tunable nonlinearity and high material quality, has become a leading choice for the realization of integrated energy-time entangled photon pairs\cite{ramirez2024integrated}.

In this work, we demonstrate an integrated time-bin-entangled photon-pair source that simultaneously achieves high efficiency and broad bandwidth through optimizing the 4H-SiC material and fabrication platform. We first optimize the effective nonlinear coefficient of the 4H-SiC material. Then we fabricate 4H-SiC microring resonator (MRR) \cite{wang2021high} with a cross-section of \( 0.9\,\mu\text{m} \times 0.4\,\mu\text{m} \). The devices operate at a loaded quality factor of \( 1.9 \times 10^5 \) and deliver a photon-pair generation rate of $1.35\times10^7$   $~\mathrm{s^{-1}mW^{-2}}$ with a spectral bandwidth of 1.0 GHz. Time-bin-entangled photon pairs are generated with a raw visibility of $95.55\pm0.18\%$, clearly violating Bell’s inequality, and quantum state tomography yields a fidelity of $94.37\pm0.22\%$. These results establish that material and device co-design can overcome the intrinsic efficiency–bandwidth trade-off in cavity-based quantum light sources, providing a scalable pathway toward integrated, high-rate entangled-photon platforms for future quantum communication networks.

\section{Device optimization and characterization}
The photon-pair generation rate in the spontaneous four-wave mixing (SFWM) process can be expressed as 
\begin{equation}
    N_{c} \propto \frac{\gamma^{2} Q^{3}}{R^{2}}\propto \frac{n_2^{2}}{A_{\text{eff}}^2R^{2}}\frac{1}{\Delta \nu^{3}}.
\end{equation}  
where $\gamma$ is the third-order nonlinear coefficient, $Q$ is the loaded quality factor, $R$ is the radius of the MRR, $n_{2}$ is the nonlinear refractive index, $A_{\text{eff}}$ is the effective mode area, $\Delta\nu$ is the linewidth of the MRR. 

For high-speed quantum communication, entangled sources must be both broadband and efficient. Broadband operation requires a large cavity linewidth $\Delta\nu$ (i.e., low $Q$), yet Eq.~(1) shows that the photon-pair rate scales as $N_c \propto \Delta\nu^{-3}$, which introduces a trade-off between the properties of broadband and efficient. We offset this penalty by maximizing the nonlinear index $n_2$, minimizing the effective mode area $A_{\mathrm{eff}}$, and reducing the ring radius $R$. 

\begin{figure}[h!]
    \centering
    \includegraphics[width=8.5 cm]{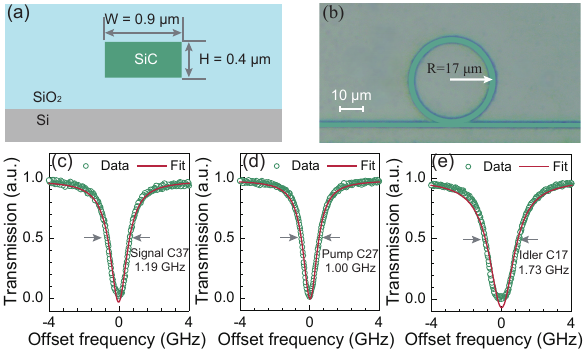}
    \caption{Device design and characterization. (a) Cross-section of 4H-SiC MRR with width-height of $0.9~\mu\mathrm{m}\times0.4~\mu\mathrm{m}$. (b) Microscopy image of the 4H-SiC MRR with a radius of 17 $\mu\mathrm{m}$. (c)-(e) Transmissions at the signal, pump, and idler channels, with a linewidth of 1.19, 1.00, 1.73 GHz, corresponding to a loaded quality factor of $1.6 \times 10^5$, $1.9 \times 10^5$, $1.1 \times 10^5$, respectively.}
    \label{fig:Fig1}
\end{figure}

To this end, we realize a 4H-SiC MRR with a waveguide cross section of $0.9~\mu\mathrm{m}\times0.4~\mu\mathrm{m}$ as shown in Fig.~\ref{fig:Fig1}(a). The material provides $n_2=8\times10^{-19}~\mathrm{m}^2\mathrm{W}^{-1}$, and the fundamental $\mathrm{TE}_{00}$ mode has $A_{\mathrm{eff}}\approx0.39~\mu\mathrm{m}^2$. As shown in Fig.~\ref{fig:Fig1}(b), the radius of the device is designed as $R=17~\mu\mathrm{m}$, corresponding to a free spectral range (FSR) of $\sim 1~\mathrm{THz}$, and a coupling gap of $0.45~\mu\mathrm{m}$ to the bus waveguide. Additional design and fabrication details are provided in the Supplemental Material Note~1 and Note~2. Figure~\ref{fig:Fig1}(c)-(e) gives the measured transmissions at signal, pump, and idler channels, i.e., ITU channel C37, C27, and C17, with a measured line width of 1.19, 1.00, and 1.73 GHz, which corresponds to a Q-factor of 1.6$\times10^5$, 1.9$\times10^5$, and 1.1$\times10^5$, respectively.

\section{Photon pair generation and characterization}
\begin{figure}[h]
    \centering
    \includegraphics[width=\linewidth]{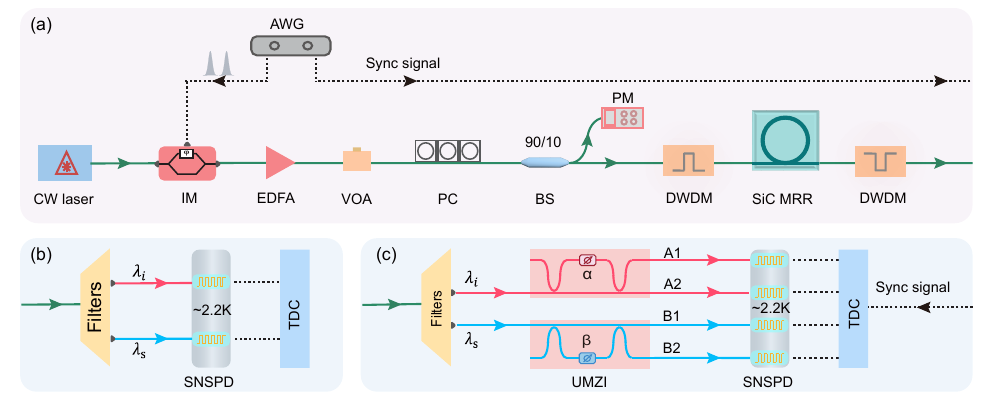}
    \caption{Experimental setup for the generation and characterization of integrated time-bin entangled photon pairs. (a) Generation of time-bin entangled photon pairs with plused-laser pump. (b) Correlation property. (c) Time-bin entanglement. CW laser: continuous wave laser, IM: intensity modulator, EDFA: erbium-doped fiber amplifier, AWG: arbitrary waveform generator, VOA: variable optical attenuator, PC: polarization controller, BS: beam splitter, PM: power meter, DWDM: dense wavelength-division multiplexer, SNSPD: superconducting nanowire single-photon detector, TDC: time-to-digital converter, UMZI: unbalanced Mach–Zehnder interferometer. $\lambda_{s}$ and $\lambda_{i}$ are wavelengths of signal and idler photons, respectively. The SNSPDs are operated at a temperature of 2.2 K.}
    \label{fig:Fig2}
\end{figure}

The experimental setup for photon pair generation and characterization is shown in Fig.~\ref{fig:Fig2}. As shown in Fig.~\ref{fig:Fig2}(a), a 4H-SiC MRR is pumped using a pulsed laser at 1555.75 nm (ITU Channel C27). The pulses are generated by modulating the continuous wave laser (CW laser) with an arbitrary waveform generator (AWG) and an intensity modulator (IM). See more details about the preparation of pulsed pump light in Supplementary Materials Note 6. The pump power and polarization are controlled via an erbium-doped fiber amplifier (EDFA), a variable optical attenuator (VOA), and a polarization controller (PC), with real-time monitoring by a power meter (PM) through a 90/10 beam splitter (BS). To minimize amplified spontaneous emission (ASE) noise from the EDFA and spontaneous Raman noise from the fiber system, two cascaded 100-GHz dense wavelength division multiplexers (DWDMs) at ITU channel 27 (C27) provide over 120 dB pump suppression. 
Light is coupled into and out of the chip through lensed fibers, with a coupling loss of 4 dB per facet. Additional cascaded DWDMs further suppress residual pump light at the output. The generated signal and idler photons are detected by superconducting nanowire single-photon detectors (SNSPDs) with a detection efficiency of 90\% and a dark count rate of 30 Hz. Photon arrival times are recorded by a time-to-digital converter (TDC) to construct coincidence histograms.

\begin{figure}[h]
    \centering
    \includegraphics[width=\linewidth]{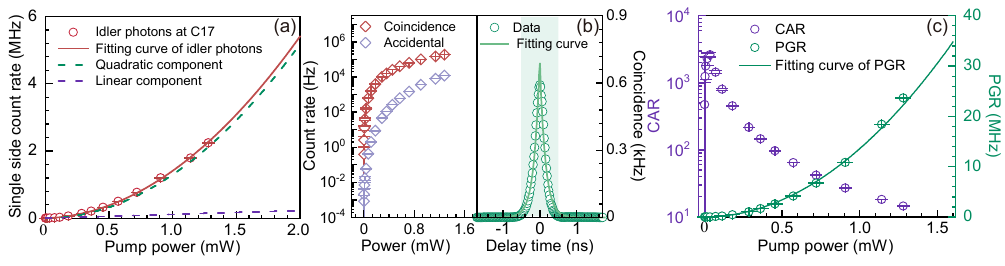}
    \caption{Correlated photon pair generation in SiC microring. (a) Single side count rate of the idler photon versus pump power. (b) The left shows the coincidence and accidental coincidence count rate versus pump power. The right is the measured coincidence histogram of the signal and idler photons at ITU-C37 and ITU-C17 under 0.29 mW. (c) Coincidence-to-accidental ratio (CAR) and PGR versus pump power.}
    \label{fig:Fig3}
\end{figure}

The generation rate of photon pairs in SFWM scales quadratically with pump power. We model the single side count rate as $R_{s/i}=a_{s/i}P^{2}+b_{s/i}P+c_{s/i}$, where $a_{s/i}$, $b_{s/i}$, and $c_{s/i}$ represent SFWM efficiency, linear noise, and dark counts, respectively, and P is the on-chip pump power\cite{wang2021integrated,engin2013photon}. In this experiment, the signal and idler photons are selected with two 100-GHz DWDMs at 1547.72 nm (ITU Channel C37) and 1563.86 nm (ITU Channel C17), respectively. Figure~\ref{fig:Fig3}(a) shows the single side count rate of idler photon as a function of on-chip pump power, reaching over 2.24 MHz at approximately 1.28 mW. A quadratic fit yields $a_i=1.29 \times 10^{6}~ \mathrm{s^{-1}mW^{-2}}$ and a linear coefficient $b_i=1.15 \times 10^{5}~s^{-1}\mathrm{mW^{-1}}$, corresponding to the green and purple dashed curves, respectively. These results confirm the high-quality generation of correlated photon pairs with low Raman noise in the 4H-SiC MRR, reflecting the high-purity crystalline quality of the material.

To further characterize the photon pairs, we measure the coincidence count rate $N_{cc}$ and accidental coincidence count rate $N_{acc}$ within a 1-ns time window, as shown in Fig.~\ref{fig:Fig3}(b). The right panel presents a coincidence histogram at 0.29 mW on-chip power, fitted with a double-exponential function exp$[-|\Delta t |/\tau]$, yielding a coherence time $\tau$ of 152 ps and a photon pair bandwidth $\Delta \nu$ of 1.0 GHz. The coincidence-to-accidental ratio (CAR) reaches a maximum of 2629 $\pm$ 207 at a coincidence count rate of 158 Hz, as shown in Fig.~\ref{fig:Fig2}(c). The photon-pair generation rate is given by $N_c={\left( a_sP^2\cdot a_iP^2 \right)}/{\left( N_{cc}-N_{acc} \right)}$, where $a_sP^2$ and $a_iP^2$ are the signal and idler counts from SFWM contributions\cite{samara2019high,moody2020chip}, with $a_s=1.19 \times 10^{6} ~\mathrm{s^{-1}mW^{-2}}$. The extracted PGR per unit on-chip pump power is 13.5 MHz, corresponding to a source brightness of $1.35\times10^{7} ~\mathrm{s^{-1}\cdot GHz^{-1} \cdot mW^{-2}}$. 

\section{Time-bin entanglement}
The time-bin entangled photons can be created when periodically repeated double pulses are used to pump the 4H-SiC microring. In our experiments, the double pulses with a duration of 300 ps, a separation of 1.25 ns, and a repetition rate of 160 MHz are modulated. The properties of the generated time-bin entanglement are characterized using the experimental setup shown in Fig.~\ref{fig:Fig2}(c). Two unbalanced Mach–Zehnder interferometers (UMZIs), each with a path-length imbalance precisely matched to the pump-pulse separation (1.25 ns), are employed to analyze the two-photon state. When coincidences are recorded between the two UMZI output ports, five distinguishable peaks appear, reflecting all possible combinations of arrival-time differences. To isolate the interference associated with the central peak, a synchronization clock is introduced, and three-fold coincidence measurements are performed among the clock and the detection events at the two UMZI output ports. The resulting quantum interference fringes follow 
 \(N_{ij}\propto1+(-1)^{(i+j)}V_{ij}\text{cos}(\alpha+\beta)\), where $i, j$ label the output ports (A1, B1, A2, and B2), and $\alpha$ and $\beta$ denote the phases of the two UMZIs.
\begin{figure}[h]
    \centering
    \includegraphics[width=\linewidth]{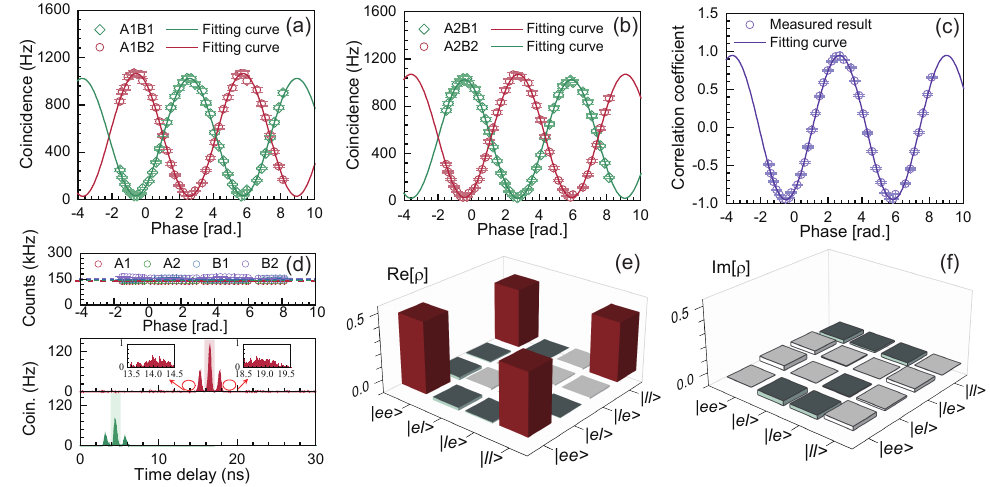}
    \caption{Characterization of time-bin entanglement. 
    (a) and (b) Interference fringes as a function of the relative phase. The coincidences are measured between the synchronous electrical signal and photon detection events at four different output port combinations: A1\&B1, A1\&B2, A2\&B1, and A2\&B2. 
    (c) Correlation coefficient. (d) Counts of the four UMZI ports and typical post-selection results based on triple coincidence. (e) Real and (f) imaginary parts of the reconstructed density matrix $\rho$ for the time-bin entangled state.}
    \label{fig:Fig4}
\end{figure}
Figure~{\ref{fig:Fig4}}(a) and (b) show two-photon interference fringes measured for all port combinations of the UMZIs(A1-B1, A1-B2, A2-B1, and A2-B2) using 1-ns coincidence windows. By fixing phase $\alpha$ in one UMZI while scanning $\beta$ in the other, we obtain post-selected interference fringes with visibilities of $ 94.63\pm0.30\%$, $95.78\pm0.42\%$, $96.99\pm0.39\%$, and $95.77\pm0.29\%$ respectively, all exceeding the CHSH violation threshold of 70.7\%. The correlation coefficient follows the expected relation $E(\alpha+\beta) = V\cos(\alpha+\beta)$, yielding a raw visibility of $95.55\pm0.18\%$ and a Bell-inequality violation by more than 138 standard deviations, as shown in Fig.~\ref{fig:Fig4}(c). The counts at the four UMZI output ports remain constant during the experiment, as shown in the top panel of Fig.~\ref{fig:Fig4}(d). The middle and bottom panels display representative two-fold coincidence counts between A1 and B1 and the corresponding three-fold coincidence counts, respectively.
Furthermore, time-bin entanglement can be verified through quantum state tomography, enabling the reconstruction of the density matrix. By implementing phase modulation and performing time-domain post-selection, the signal and idler photons are projected onto 16 combinations of the four basis states $\{|e\rangle, |l\rangle, |+\rangle, |+i\rangle\}$, where $|+\rangle=(|e\rangle+|l\rangle)/\sqrt{2}$, $|+i\rangle=(|e\rangle+i|l\rangle)/\sqrt{2}$. Coincidence measurements in these bases allow for high-fidelity reconstruction of the quantum state's density matrix. The experimentally reconstructed density matrix is given by
 
\begin{equation}
\small
\rho =
\begin{bmatrix}
 0.5300 & -0.0215 + i\,0.0355 & -0.0089 + i\,0.0296 & 0.4530 - i\,0.0292 \\
 -0.0215 - i\,0.0355 & 0.0124 & 0.0078 + i\,0.0056 & -0.0038 - i\,0.0126 \\
 -0.0089 - i\,0.0296 & 0.0078 - i\,0.0056 & 0.0092 & 0.0116 - i\,0.0260 \\
 0.4530 + i\,0.0292 & -0.0038 + i\,0.0126 & 0.0116 + i\,0.0260 & 0.4470
\end{bmatrix}.
\end{equation}

The real and imaginary components of $\rho$ are depicted in Fig.~\ref{fig:Fig4}(e) and (f), respectively. Relative to the maximally entangled Bell state $|\phi^+\rangle = (|ee\rangle + |ll\rangle)/\sqrt{2}$, the fidelity of the reconstructed state is $94.37\pm0.22\%$, with uncertainty estimated via 1000-time Monte Carlo simulations. This result confirms that the experimentally prepared entangled state closely approximates the ideal maximally entangled state, demonstrating the high coherence and purity of the generated time-bin entanglement.

\section{Discussion and conclusion}
Our results demonstrate that co-engineering the nonlinear optical properties of 4H-SiC and the geometry of a microcavity offers a powerful approach to overcoming the long-standing efficiency–bandwidth trade-off in cavity-enhanced photon-pair sources. By exploiting the material-engineered nonlinearity, sub-micron mode confinement, and a small-radius microresonator, we have achieved a 1-GHz cavity bandwidth and a photon-pair generation rate exceeding $1.35\times10^{7}~\mathrm{s}^{-1}\,\mathrm{mW}^{-2}$. Despite the trade-off associated with a lower quality factor $Q$, the PGR surpasses previous demonstrations on similar platforms with a higher $Q$. This improvement is primarily driven by an increase in the non-linear refractive index $n_2$, a reduction in the effective mode area $A_{\text{eff}}$ and a decrease in the ring radius \( R \) in our experiments. Further details can be found in the Supplemental Material Note 2. The interplay between $Q$ and $\gamma$ represents a fundamental trade-off in integrated quantum photonics. While a high $Q$ factor is traditionally favored in SFWM due to reduced propagation losses, our findings indicate that enhancing $\gamma$ can compensate for efficiency losses caused by lower $Q$. These optimization strategies extend beyond 4H-SiC to other platforms such as Si, SiN, and AlN. With such a quantum light source, high-speed quantum communication, such as quantum teleportation and quantum key distribution, could benefit significantly, enabling improved rates and enhanced scalability for future quantum networks.

In summary, we have demonstrated an integrated time-bin entangled source based on a 4H-SiC MRR, achieving efficient photon-pair generation via nonlinear coefficient engineering. The source exhibits a PGR of $1.35\times10^7$   $~\mathrm{s^{-1} mW^{-2}}$ and a time-bin entanglement fidelity of $94.37\pm0.22\%$. Our results highlight the trade-off between nonlinear enhancement and cavity \( Q \)-factors, demonstrating that increased nonlinear coefficients can compensate for efficiency losses associated with lower \( Q \), enabling short coherence times crucial for high-repetition-rate photon generation. This approach provides a scalable pathway for integrated quantum photonic devices and high-speed quantum networks.

\begin{backmatter}
\bmsection{Funding}
This work was supported by the National Key R\&D Program of China (No.~2022YFA1404600); the Sichuan Science and Technology Program (Nos.~2024YFHZ0369, 2024YFHZ0370, 2024YFHZ0368, 2026NSFSC1439); the National Natural Science Foundation of China (Nos.~62475039, 62405046, 62375043, 62293521, 12575313, 12074400, and 62205363); the CAS Project for Young Scientists in Basic Research (No.~YSBR-69); and the Quantum Science and Technology—National Science and Technology Major Project (Nos.~2024ZD0300800 and 2021ZD0300701).

\bmsection{Acknowledgment}

\bmsection{Disclosures}
The authors declare no conflicts of interest.

\bmsection{Data Availability Statement}
All of the data that support the findings of this study are
reported in the maintext and Supplement 1. Source data are available from the corresponding authors on reasonable request.

\bmsection{Supplemental document}
See Supplement 1 for supporting content.

\end{backmatter}



\begin{thebibliography}{10}
\newcommand{\enquote}[1]{``#1''}

\bibitem{marcikic2003long}
I.~Marcikic, H.~De~Riedmatten, W.~Tittel, \emph{et~al.}, \enquote{Long-distance teleportation of qubits at telecommunication wavelengths,} {\protect\JournalTitle{Nature}} \textbf{421}, 509--513 (2003).

\bibitem{zhang2025classical}
Y.~Zhang, R.~Broberg, A.~Zhu, \emph{et~al.}, \enquote{Classical-decisive quantum internet by integrated photonics,} {\protect\JournalTitle{Science}} \textbf{389}, 940--944 (2025).

\bibitem{gottesman1999demonstrating}
D.~Gottesman and I.~L. Chuang, \enquote{Demonstrating the viability of universal quantum computation using teleportation and single-qubit operations,} {\protect\JournalTitle{Nature}} \textbf{402}, 390--393 (1999).

\bibitem{psiquantum2025manufacturable}
P.~team, \enquote{A manufacturable platform for photonic quantum computing,} {\protect\JournalTitle{Nature}} \textbf{641}, 876--883 (2025).

\bibitem{aghaee2025scaling}
H.~Aghaee~Rad, T.~Ainsworth, R.~Alexander, \emph{et~al.}, \enquote{Scaling and networking a modular photonic quantum computer,} {\protect\JournalTitle{Nature}} \textbf{638}, 912--919 (2025).

\bibitem{afek2010high}
I.~Afek, O.~Ambar, and Y.~Silberberg, \enquote{High-noon states by mixing quantum and classical light,} {\protect\JournalTitle{Science}} \textbf{328}, 879--881 (2010).

\bibitem{cao2024multi}
A.~Cao, W.~J. Eckner, T.~Lukin~Yelin, \emph{et~al.}, \enquote{Multi-qubit gates and schr{\"o}dinger cat states in an optical clock,} {\protect\JournalTitle{Nature}} \textbf{634}, 315--320 (2024).

\bibitem{finkelstein2024universal}
R.~Finkelstein, R.~B.-S. Tsai, X.~Sun, \emph{et~al.}, \enquote{Universal quantum operations and ancilla-based read-out for tweezer clocks,} {\protect\JournalTitle{Nature}} \textbf{634}, 321--327 (2024).

\bibitem{wang2018multidimensional}
J.~Wang, S.~Paesani, Y.~Ding, \emph{et~al.}, \enquote{Multidimensional quantum entanglement with large-scale integrated optics,} {\protect\JournalTitle{Science}} \textbf{360}, 285--291 (2018).

\bibitem{kwiat2001experimental}
P.~G. Kwiat, S.~Barraza-Lopez, A.~Stefanov, and N.~Gisin, \enquote{Experimental entanglement distillation and ‘hidden’non-locality,} {\protect\JournalTitle{Nature}} \textbf{409}, 1014--1017 (2001).

\bibitem{erhard2018twisted}
M.~Erhard, R.~Fickler, M.~Krenn, and A.~Zeilinger, \enquote{Twisted photons: new quantum perspectives in high dimensions,} {\protect\JournalTitle{Light: Science \& Applications}} \textbf{7}, 17146--17146 (2018).

\bibitem{krenn2014generation}
M.~Krenn, M.~Huber, R.~Fickler, \emph{et~al.}, \enquote{Generation and confirmation of a (100$\times$ 100)-dimensional entangled quantum system,} {\protect\JournalTitle{Proceedings of the National Academy of Sciences}} \textbf{111}, 6243--6247 (2014).

\bibitem{kues2017chip}
M.~Kues, C.~Reimer, P.~Roztocki, \emph{et~al.}, \enquote{On-chip generation of high-dimensional entangled quantum states and their coherent control,} {\protect\JournalTitle{Nature}} \textbf{546}, 622--626 (2017).

\bibitem{franson1989bell}
J.~D. Franson, \enquote{Bell inequality for position and time,} {\protect\JournalTitle{Physical review letters}} \textbf{62}, 2205 (1989).

\bibitem{marcikic2002time}
I.~Marcikic, H.~de~Riedmatten, W.~Tittel, \emph{et~al.}, \enquote{Time-bin entangled qubits for quantum communication created by femtosecond pulses,} {\protect\JournalTitle{Physical Review A}} \textbf{66}, 062308 (2002).

\bibitem{marcikic2004distribution}
I.~Marcikic, H.~De~Riedmatten, W.~Tittel, \emph{et~al.}, \enquote{Distribution of time-bin entangled qubits over 50 km of optical fiber,} {\protect\JournalTitle{Physical review letters}} \textbf{93}, 180502 (2004).

\bibitem{reimer2016generation}
C.~Reimer, M.~Kues, P.~Roztocki, \emph{et~al.}, \enquote{Generation of multiphoton entangled quantum states by means of integrated frequency combs,} {\protect\JournalTitle{Science}} \textbf{351}, 1176--1180 (2016).

\bibitem{yu2025quantum}
H.~Yu, S.~Sciara, M.~Chemnitz, \emph{et~al.}, \enquote{Quantum key distribution implemented with d-level time-bin entangled photons,} {\protect\JournalTitle{Nature Communications}} \textbf{16}, 171 (2025).

\bibitem{yu2024time}
H.~Yu, A.~O. Govorov, H.-Z. Song, and Z.~Wang, \enquote{Time-encoded photonic quantum states: Generation, processing, and applications,} {\protect\JournalTitle{Applied Physics Reviews}} \textbf{11} (2024).

\bibitem{singh2025photonic}
A.~Singh, A.~Sethia, L.~Esmaeilifar, \emph{et~al.}, \enquote{Photonic quantum information with time-bins: Principles and applications,} {\protect\JournalTitle{arXiv preprint arXiv:2507.08102}}  (2025).

\bibitem{xiong2015compact}
C.~Xiong, X.~Zhang, A.~Mahendra, \emph{et~al.}, \enquote{Compact and reconfigurable silicon nitride time-bin entanglement circuit,} {\protect\JournalTitle{Optica}} \textbf{2}, 724--727 (2015).

\bibitem{zhang2018integrated}
X.~Zhang, B.~A. Bell, A.~Mahendra, \emph{et~al.}, \enquote{Integrated silicon nitride time-bin entanglement circuits,} {\protect\JournalTitle{Optics letters}} \textbf{43}, 3469--3472 (2018).

\bibitem{caspani2017integrated}
L.~Caspani, C.~Xiong, B.~J. Eggleton, \emph{et~al.}, \enquote{Integrated sources of photon quantum states based on nonlinear optics,} {\protect\JournalTitle{Light: Science \& Applications}} \textbf{6}, e17100--e17100 (2017).

\bibitem{wang2020integrated}
J.~Wang, F.~Sciarrino, A.~Laing, and M.~G. Thompson, \enquote{Integrated photonic quantum technologies,} {\protect\JournalTitle{Nature Photonics}} \textbf{14}, 273--284 (2020).

\bibitem{elshaari2020hybrid}
A.~W. Elshaari, W.~Pernice, K.~Srinivasan, \emph{et~al.}, \enquote{Hybrid integrated quantum photonic circuits,} {\protect\JournalTitle{Nature photonics}} \textbf{14}, 285--298 (2020).

\bibitem{mahmudlu2023fully}
H.~Mahmudlu, R.~Johanning, A.~Van~Rees, \emph{et~al.}, \enquote{Fully on-chip photonic turnkey quantum source for entangled qubit/qudit state generation,} {\protect\JournalTitle{Nature Photonics}} \textbf{17}, 518--524 (2023).

\bibitem{wang2025scalable}
H.~Wang, T.~C. Ralph, J.~J. Renema, \emph{et~al.}, \enquote{Scalable photonic quantum technologies,} {\protect\JournalTitle{Nature Materials}} pp. 1--15 (2025).

\bibitem{kramnik2025scalable}
D.~Kramnik, I.~Wang, A.~Ramesh, \emph{et~al.}, \enquote{Scalable feedback stabilization of quantum light sources on a cmos chip,} {\protect\JournalTitle{Nature Electronics}} pp. 1--11 (2025).

\bibitem{guo2017parametric}
X.~Guo, C.-l. Zou, C.~Schuck, \emph{et~al.}, \enquote{Parametric down-conversion photon-pair source on a nanophotonic chip,} {\protect\JournalTitle{Light: Science \& Applications}} \textbf{6}, e16249--e16249 (2017).

\bibitem{reimer2019high}
C.~Reimer, S.~Sciara, P.~Roztocki, \emph{et~al.}, \enquote{High-dimensional one-way quantum processing implemented on d-level cluster states,} {\protect\JournalTitle{Nature Physics}} \textbf{15}, 148--153 (2019).

\bibitem{ma2020ultrabright}
Z.~Ma, J.-Y. Chen, Z.~Li, \emph{et~al.}, \enquote{Ultrabright quantum photon sources on chip,} {\protect\JournalTitle{Physical Review Letters}} \textbf{125}, 263602 (2020).

\bibitem{steiner2021ultrabright}
T.~J. Steiner, J.~E. Castro, L.~Chang, \emph{et~al.}, \enquote{Ultrabright entangled-photon-pair generation from an al ga as-on-insulator microring resonator,} {\protect\JournalTitle{Prx Quantum}} \textbf{2}, 010337 (2021).

\bibitem{fan2023multi}
Y.-R. Fan, C.~Lyu, C.-Z. Yuan, \emph{et~al.}, \enquote{Multi-wavelength quantum light sources on silicon nitride micro-ring chip,} {\protect\JournalTitle{Laser \& Photonics Reviews}} \textbf{17}, 2300172 (2023).

\bibitem{chen2024ultralow}
R.~Chen, Y.-H. Luo, J.~Long, \emph{et~al.}, \enquote{Ultralow-loss integrated photonics enables bright, narrowband, photon-pair sources,} {\protect\JournalTitle{Physical Review Letters}} \textbf{133}, 083803 (2024).

\bibitem{zeng2024quantum}
H.~Zeng, Z.-Q. He, Y.-R. Fan, \emph{et~al.}, \enquote{Quantum light generation based on gan microring toward fully on-chip source,} {\protect\JournalTitle{Physical Review Letters}} \textbf{132}, 133603 (2024).

\bibitem{rahmouni2024entangled}
A.~Rahmouni, R.~Wang, J.~Li, \emph{et~al.}, \enquote{Entangled photon pair generation in an integrated sic platform,} {\protect\JournalTitle{Light: Science \& Applications}} \textbf{13}, 110 (2024).

\bibitem{pang2025versatile}
Y.~Pang, J.~E. Castro, T.~J. Steiner, \emph{et~al.}, \enquote{Versatile chip-scale platform for high-rate entanglement generation using an al ga as microresonator array,} {\protect\JournalTitle{PRX Quantum}} \textbf{6}, 010338 (2025).

\bibitem{Luo17QiangL}
R.~Luo, H.~Jiang, S.~Rogers, \emph{et~al.}, \enquote{On-chip second-harmonic generation and broadband parametric down-conversion in a lithium niobate microresonator,} {\protect\JournalTitle{Opt. Express}} \textbf{25}, 24531--24539 (2017).

\bibitem{Kumar20}
R.~R. Kumar, Y.~Wang, Y.~Zhang, and H.~K. Tsang, \enquote{Quantum states of higher-order whispering gallery modes in a silicon micro-disk resonator,} {\protect\JournalTitle{Journal of the Optical Society of America B}} \textbf{37}, 2231--2237 (2020).

\bibitem{Ren2022microdisk}
B.-Y. Xu, L.-K. Chen, J.-T. Lin, \emph{et~al.}, \enquote{Spectrally multiplexed and bright entangled photon pairs in a lithium niobate microresonator,} {\protect\JournalTitle{Science China Physics, Mechanics \& Astronomy}} \textbf{65}, 294262 (2022).

\bibitem{Ortiz-Ricardo2021}
E.~O. Ricardo, C.~Bertoni-Ocampo, M.~Maldonado-Terrón, \emph{et~al.}, \enquote{Submegahertz spectral width photon-pair source based on fused silica microspheres,} {\protect\JournalTitle{Photonics Research}} \textbf{9}, 110 (2021).

\bibitem{ramelow2015silicon}
S.~Ramelow, A.~Farsi, S.~Clemmen, \emph{et~al.}, \enquote{Silicon-nitride platform for narrowband entangled photon generation,} {\protect\JournalTitle{arXiv preprint arXiv:1508.04358}}  (2015).

\bibitem{samara2019high}
F.~Samara, A.~Martin, C.~Autebert, \emph{et~al.}, \enquote{High-rate photon pairs and sequential time-bin entanglement with si3n4 microring resonators,} {\protect\JournalTitle{Optics express}} \textbf{27}, 19309--19318 (2019).

\bibitem{xu2025quantum}
N.~Chen, W.-X. Li, Y.-R. Fan, \emph{et~al.}, \enquote{Quantum light sources with configurable lifetime leveraging parity-time symmetry,} {\protect\JournalTitle{arXiv preprint arXiv:2504.01413}}  (2025).

\bibitem{chen2023pushing}
N.~Chen, Z.~Wang, J.~Wu, \emph{et~al.}, \enquote{Pushing photon-pair generation rate in microresonators by q factor manipulation,} {\protect\JournalTitle{Optics Letters}} \textbf{48}, 5355--5358 (2023).

\bibitem{li2023measurement}
J.~Li, R.~Wang, L.~Cai, and Q.~Li, \enquote{Measurement of the kerr nonlinear refractive index and its variation among 4 h-si c wafers,} {\protect\JournalTitle{Physical Review Applied}} \textbf{19}, 034083 (2023).

\bibitem{lu2024strong}
Y.~Lu, X.~Shi, A.~Ali~Afridi, \emph{et~al.}, \enquote{Strong third-order nonlinearity in amorphous silicon carbide waveguides,} {\protect\JournalTitle{Optics Letters}} \textbf{49}, 4389--4392 (2024).

\bibitem{cai2022octave}
L.~Cai, J.~Li, R.~Wang, and Q.~Li, \enquote{Octave-spanning microcomb generation in 4h-silicon-carbide-on-insulator photonics platform,} {\protect\JournalTitle{Photonics Research}} \textbf{10}, 870--876 (2022).

\bibitem{ramirez2024integrated}
P.~T. Ram{\'\i}rez, J.~D. G{\'o}mez, G.~R. Becerra, \emph{et~al.}, \enquote{Integrated photon pairs source in silicon carbide based on micro-ring resonators for quantum storage at telecom wavelengths,} {\protect\JournalTitle{Scientific Reports}} \textbf{14}, 17755 (2024).

\bibitem{wang2021high}
C.~Wang, Z.~Fang, A.~Yi, \emph{et~al.}, \enquote{High-q microresonators on 4h-silicon-carbide-on-insulator platform for nonlinear photonics,} {\protect\JournalTitle{Light: Science \& Applications}} \textbf{10}, 139 (2021).

\bibitem{wang2021integrated}
Y.~Wang, K.~D. J{\"o}ns, and Z.~Sun, \enquote{Integrated photon-pair sources with nonlinear optics,} {\protect\JournalTitle{Applied Physics Reviews}} \textbf{8} (2021).

\bibitem{engin2013photon}
E.~Engin, D.~Bonneau, C.~M. Natarajan, \emph{et~al.}, \enquote{Photon pair generation in a silicon micro-ring resonator with reverse bias enhancement,} {\protect\JournalTitle{Optics express}} \textbf{21}, 27826--27834 (2013).

\bibitem{moody2020chip}
G.~Moody, L.~Chang, T.~J. Steiner, and J.~E. Bowers, \enquote{Chip-scale nonlinear photonics for quantum light generation,} {\protect\JournalTitle{AVS Quantum Science}} \textbf{2} (2020).

\end{thebibliography}

\end{document}